\documentclass{easychair}
\usepackage{amsmath}
\usepackage{amssymb}
\usepackage{lstsemantic}
\usepackage{lstlangcoq}
\usepackage{xspace}
\usepackage{booktabs}
\newtheorem{definition}{Definition}

\def\libname#1{\textsf{#1}\xspace}
\def\systemname#1{\textsf{#1}\xspace}
\newcommand{\coq}{\systemname{Coq}}
\newcommand{\tmegg}{\systemname{tmEgg}}
\newcommand{\corn}{\libname{CoRN}}

\begin{document}
\title{Machine Learning of \coq Proof Guidance: \\
        First Experiments}
\titlerunning{Machine Learning of \coq Proof Guidance}
\author{
     Cezary Kaliszyk\inst{1} %
\and Lionel Mamane\inst{2}
\and Josef Urban\inst{3}
}

\institute{
  University of Innsbruck,
  Austria\\
  \email{cezary.kaliszyk@uibk.ac.at}
\and
  Luxembourg\\
  \email{lionel@mamane.lu}\\
\and
  Radboud Uniersity Nijmegen,
  Netherlands\\
  \email{josef.urban@gmail.com}\\
 }
\authorrunning{Kaliszyk, Mamane, and Urban}

\clearpage
\maketitle

\begin{abstract}
  We report the results of the first experiments with learning proof
  dependencies from the formalizations done with the \coq system. We
  explain the process of obtaining the dependencies from the \coq
  proofs, the characterization of formulas that is used for the
  learning, and the evaluation method. Various machine learning
  methods are compared on a dataset of 5021 toplevel \coq proofs coming
  from the \corn repository. The best resulting method covers on
  average 75\% of the needed proof dependencies among the first 100
  predictions, which is a comparable performance of such initial
  experiments on other large-theory corpora.
\end{abstract}

\section{Introduction}\label{sec:intro}

In the last decade, a number of bridges between interactive theorem
provers (ITPs) and various external proof-advising systems have been
built. The best-known examples of such systems are today the bridges
between ITPs such as Isabelle, Mizar and HOL Light and automated
theorem provers (ATPs) such as Vampire, E, and Z3. The success rate of
such bridges for Mizar and HOL Light has recently reached 40\% in a
scenario that assumes no assistance from the users, only the ability
to learn from the previous proofs that are already available in the
large libraries of these ITP systems. Important components of these
systems are the premise-selection algorithms, i.e., algorithms that
for a given conjecture choose the most relevant theorems and
definitions that are already available in the large libraries. Such
algorithms can be actually used separately as a search engine
available to the ITP users in cases when the full automated proof is
too hard to find. Such functionality has been implemented early for
the Mizar system and integrated in its Emacs authoring environment in
the form of the Mizar Proof Advisor~\cite{Urban2006414}. Good premise
selection is thus a prerequisite for the fully automated AI/ATP
systems, and also a functionality that can be immediately given to the
ITP users.

In this work, we develop premise selection techniques for the \coq
proof assistant and evaluate their performance on the Constructive \coq
Repository at Nijmegen (\corn)~\cite{Cruz-FilipeGW04}, containing 5021
toplevel proofs of theorems such as the Fundamental Theorem of
Algebra~\cite{GWZ02}. The necessary work involves extraction of proof
dependencies, for which we re-use previous work targeted at \coq user
interfaces by Mamane, see Section~\ref{sec:deps}. In the
type-theoretical context used by \coq there is no clear distinction
between theorems and definitions on one side, and formula symbols and
terms on the other side. This makes it difficult to see from the
extracted dependency data which of the dependencies characterize the
conjecture to be proved, and which of them are the ``proper'' proof
dependencies that we would like to recommend to the user (and
ultimately to an ATP system). Section~\ref{sec:syms} explains how we
heuristically differentiate between these two kinds of
data. Section~\ref{sec:learn} briefly explains how we learn premise
selection from such data, and Section~\ref{sec:results} presents the
first experimental results. It is shown  that the best resulting method covers on
  average 75\% of the needed proof dependencies among the first 100
  predictions, which is a comparable performance of such initial
  experiments on other large-theory corpora. Section~\ref{sec:concl} discusses future work and concludes.

\section{Extracting \coq Dependencies}\label{sec:deps}
The processs of extracting proof dependencies from \coq developments that we use
in this work is described in detail in~\cite{AlamaMU12}. Here we give a brief overview of this process and related issues.

\coq is based
on the Curry-Howard isomorphism, which means that
statements (formulas) are encoded as types, 
  and at the logical level there is
  no essential difference between
  a definition and a theorem:
  they are both just the binding (in the environment)
  of a name to a type
  (type of the definition, statement of the theorem)
  and a term
  (body of the definition, proof of the theorem).
Similarly,
  there is
  no essential difference
  between
  an axiom and a parameter:
  they are both just the binding (in the environment)
  of a name to a type
  (statement of the axiom,
  type of the parameter, e.g. ``natural number'').

As part of the development of \tmegg \cite{LEM_tmegg},
Mamane added to \coq a partial facility
to communicate proof dependencies to the user interface.
This %
facility has been already integrated
in the official release of \coq.
Since then this facility was extended to be able to treat the whole of
the \corn library, however these changes are not yet included in the official release of \coq.
This facility works by tracking dependencies for various \coq commands.
There are essentially three groups of \coq commands
that need to be treated:
\begin{enumerate}
\item Commands that register a new logical construct
  (definition or axiom),
  either
  \begin{itemize}
  \item From scratch.
    That is, commands that take as arguments
    a name and a type and/or a body,
    and that add
    the definition binding this name
    to this type and/or body.
    The canonical examples are
    \begin{lstlisting}[language=Coq]
Definition Name : type := body
    \end{lstlisting}

and

\begin{lstlisting}[language=Coq]
Axiom Name : type
\end{lstlisting}

    The type can also be given implicitly
    as the inferred type of the body,
    as in
    \begin{lstlisting}[language=Coq]
Definition Name := body
    \end{lstlisting}

  \item Saving the current (completely proven) theorem
    in the environment.
    These are
    the ``end of proof'' commands,
    such as \texttt{Qed}, \texttt{Save}, \texttt{Defined}.
  \end{itemize}
\item Commands that make progress
  in the current proof,
  which is necessarily made in several steps:
  \begin{enumerate}
  \item Opening a new theorem,
    as in
    \lstset{numbers=none}\begin{lstlisting}[language=Coq]
Theorem Name : type
    \end{lstlisting}

    or
    \begin{lstlisting}[language=Coq]
Definition Name : type
    \end{lstlisting}

  \item An arbitrary strictly positive
    amount of proof steps.
  \item Saving that theorem in the environment.
  \end{enumerate}
  These commands update
  (by adding exactly \emph{one} node)
  the internal \coq structure called
  ``proof tree''.
\item Commands that open a new theorem,
  that will be proven in multiple steps.
\end{enumerate}
The dependency tracking
is implemented as suitable hooks in the \coq functions
that the three kinds of commands eventually call.
When a new construct is registered in the environment,
the dependency tracking
walks over
the type
and body (if present)
of the new construct
and collects all constructs that are referenced.
When a  proof tree is updated,
the dependency tracking
examines
the top node
of the new proof tree
(note that this is always
the only change
with regards to the previous proof tree).
The commands
that update the proof tree
(that is, make a step in the current proof)
are called \texttt{tactics}.
\coq's tactic interpretation goes through three main phases:
\begin{enumerate}
\item parsing;
\item Ltac\footnote{Ltac is the \coq's tactical language,
used to combine tactics and add new user-defined tactics.} expansion;
\item evaluation.
\end{enumerate}
The tactic structure after each of these phases
is stored in the proof tree.
This allows to collect all construct references
mentioned at any of these tree levels.
For example, if tactic \texttt{Foo T} is defined as
\lstset{numbers=none}\begin{lstlisting}[language=Coq]
try apply BolzanoWeierstrass;
solve [ T | auto ]
\end{lstlisting}
and the user invokes the tactic as \texttt{Foo FeitThompson},
then the first level will contain (in parsed form)
\texttt{Foo FeitThompson},
the second level will contain (in parsed form)
\begin{lstlisting}[language=Coq]
try apply BolzanoWeierstrass;
solve [ FeitThompson | auto ].
\end{lstlisting}
and the third level can contain any of:
\begin{itemize}
\item \texttt{refine (BolzanoWeierstrass \dots)},
\item \texttt{refine (FeitThompson \dots)},
\item something else, if the proof was found by \texttt{auto}.
\end{itemize}
The third level typically contains only
a few of the basic atomic fundamental rules (tactics) applications,
such as \texttt{refine}, \texttt{intro}, \texttt{rename} or \texttt{convert},
and combinations thereof.

Dependency tracking is available
in the program implementing
the \texttt{coq-interface} protocol which is designed for machine interaction between \coq and its user interfaces.
The dependency information is printed
in a special message
for each progress-making \coq command
that can give rise to a dependency. After some postprocessing, the dependency data obtained from such commands will look as follows 
(first comes the defined/proved construct, then a list of its dependencies):
\begin{verbatim}
"CoRN.algebra.Basics.NEG_anti_convert" 
  ("Coq.Init.Datatypes.S" "Coq.Init.Datatypes.nat" "Coq.Init.Datatypes.nat_ind" 
   "Coq.Init.Logic.eq" "Coq.Init.Logic.refl_equal" "Coq.NArith.BinPos.P_of_succ_nat"
   "Coq.NArith.BinPos.Psucc" "Coq.NArith.BinPos.xH" "Coq.ZArith.BinInt.Z"
   "Coq.ZArith.Bi nInt.Z_of_nat" "Coq.ZArith.BinInt.Zneg" "Coq.ZArith.BinInt.Zopp")
\end{verbatim}

\section{Learning Data}\label{sec:syms}

In order to use machine learning for guessing proof dependencies, two sets
of data need to be extracted from a proof assistant library: suitable
features of the theorems and their proof dependencies.
Machine learning for premise selection (i.e., for guessing the necessary proof dependencies) has so far been successfully
applied and evaluated for systems based on set theory
(MizAR~\cite{abs-1109-0616}) and higher order logic (Sledgehammer:MaSh~\cite{KuhlweinBKU13},
HOL(y)Hammer~\cite{holyhammer,hhmcs}).  In both kinds of systems, the
available concepts are clearly divided in different categories by the
implementation: In HOL-based systems types, type-classes and function symbols
are all distinct from theorems. In particular, theorems can not
appear inside terms. Similarly in Mizar, the meta-logic functions,
predicates, modes, and registrations are separate from the defined
set-theory objects.

This is no longer true in type-theory based systems which 
implement the Curry-Howard correspondence treating propositions as types, and proofs as terms of such (proposition) types.
Each time a named theorem $T$ is used in a proof of theorem $S$, we are actually using (a reference to) the (named) proof term of $T$. Such 
named proof terms are in \coq treated the same way as any other defined constants.
It is not possible to
distinguish named theorems from other defined constants without checking whether
the type of the constant is of a propositional type.
This is however not sufficient for us: the user may state (as a goal) and prove the
existence of values in any type (such as \texttt{nat->nat}), and premise selection should also help
in such non-propositional proofs. The users may even define their own
proposition type (different from the default type \texttt{Prop}) 
 and this is actually done in some \coq developments, for
example in the case of the \texttt{CProp} used pervasively in \corn.

The most straightforward way of adapting the machine learning
infrastructure to \coq, could be as follows. For each defined constant
$T$, the features $F(T)$ of $T$ would be the set of defined constants
present in the type (statement) of $T$, and the dependencies $D(T)$ of
$T$ would be the set of defined constants used in the proof term of $T$.
This has however one obvious disadvantage: all theorems
that talk about natural numbers would include \texttt{nat}, \texttt{O},
and \texttt{Suc} as their dependencies. Learning such constants would
also predict them, however they are currently useless in automated
techniques. To avoid learning such constants as dependencies, 
we first globally (for a given corpus $C$, such as \corn ) partition the set 
of all defined constants into those that may be used as features ($F_C$), and those that may be used as dependencies ($D_C$),
 using the following heuristic. 
 The set $F_C$ (allowed features) will consist of all the defined constants
 that appear in the types of theorems (and definitions, etc.) of $C$. 
 The set $D_C$ (allowed dependencies) will consist of all the defined constants
 that appear in all the proof terms, minus the set $F_C$.

Given a named theorem $T$, $F(T)$ is then defined as the elements of $F_C$
present in the type (statement) of $T$, while $D(T)$ are the elements of $D_C$ used in the proof term of $T$.
This
heuristic might not work well in the case where necessary propositional
constants also appear in the types (statements) of theorems; however it works
well with most \coq developments, in particular we have manually checked (on a random subset) that it behaves well with the \corn development,
which  we focus on in this paper.

Table~\ref{Stats1} shows some statistics for the machine learning data obtained.
\begin{table}
\begin{tabular}{lc}\toprule
  Accessible part of \coq Standard Library: & 5099 available facts \\
  Evaluated \corn proofs & 5021 defined (proved) theorems\\
  Distinct features & 2683\\
  Average number of features in a \corn theorem & 9.82 \\
  Average number of dependencies for a \corn theorem & 11.27\\\bottomrule
\end{tabular}
\caption{Statistics for the machine learning data}
\label{Stats1}
\end{table}
These are comparable to the data for the initial experiments done for
other systems~\cite{KaliszykU12,JoostenKU14}. Note that we do not yet work
with more advanced features such as terms, subterms, patterns, etc.

\section{Machine Learning}\label{sec:learn}

In the evaluations we use our custom implementations of three machine learning
algorithms and their combinations. These are k-Nearest Neighbours~\cite{EasyChair:74},
sparse Naive Bayes~\cite{KuhlweinBKU13}, and our modified version of the Meng-Paulson (MePo)
relevance filter~\cite{MengP09}. In order to combine the results of the algorithms
we consider the Ensemble methods~\cite{Polikar06}, in particular the weighted harmonic
average of the classifications~\cite{KaliszykU13b}. All these methods are implemented as fast tools 
(written either in OCaml or in C++) 
that are external to the \coq implementation.

\paragraph{k-Nearest Neighbours:}
The (distance-weighted) k-Nearest Neighbours learning algorithm first
finds a fixed number ($k$) of proved facts nearest (in its feature
representation) to the conjecture $c$, and then weights the
dependencies of each such fact $f$ by the distance between $f$ and
$c$. The final ranking (relevance) of dependencies is obtained by
summing their weights across the $k$ nearest neighbors.

\paragraph{Naive Bayes:}
Naive Bayes is a statistical learning method based on Bayes theorem with a strong (or naive) independence assumption.
Given a conjecture $c$ and a fact $f$, naive Bayes computes the
probability of $f$ being needed to prove $c$, based on the previous use of $f$ in proving conjectures that are
similar to $c$. The similarity is in our case expressed
using the features $F$ of the formulas.
The independence assumption says that the (non-)occurrence of a feature is not related to the (non-)occurrence of every other features.
The  predicted relevance of a fact $f$
with the set of features $F(f)$ is estimated by the conditional probability
$P(f \textrm{ is relevant} | F(c) )$.

\paragraph{MePo:}
The MePo (\emph{Me}ng--\emph{P}auls\emph{o}n) 
filter keeps track of a set of \emph{relevant features} 
initially consisting of all the conjecture's features. %
It performs the following steps iteratively, until
$n$~facts have been selected:
(i) Compute each fact's score, as roughly given by $r /
(r + i)$, where $r$ is the number of relevant features and $i$ the number
of irrelevant features occurring in the fact.
(ii) Select all facts with perfect scores as well as some of the remaining
top-scoring facts, and add all their features to the set of relevant features.

\paragraph{} Each of the above algorithms orders the available (accessible) premises w.r.t. the
likelihood that it is useful in proving the goal. In order to evaluate such
orderings, we compute the first 1024 advised premises and compute five machine
learning statistics for each of them: cover, precision, full recall, AUC, and
the average rank. Below we provide the definitions of these concepts:

\begin{definition}[100Cover]
The average coverage of the set of proof dependencies by the first 100 
suggestions is called the
(100-)Cover; it is usually expressed as a percentage of the covered proof dependencies.
\end{definition}
\begin{definition}[100Precision]
The ratio of the correctly predicted  proof dependencies in the first 100 suggestions  
is called (100-)Precision.
\end{definition}
\begin{definition}[Recall]
  Full Recall (also called 100\% Recall) is the minimum number of predictions needed to
  include the whole set of proof dependencies; or the number of predictions plus 1 if the considered
  predictions set is not large enough.
\end{definition}
\begin{definition}[Rank]
  (Average) Rank is the average position of a proof dependency in the sequence of suggestions ordered by their expected relevance.
\end{definition}

The AUC (Area under the ROC Curve) is the probability that, given a randomly drawn used premise and a randomly drawn unused premise, the used premise is ranked higher than the unused premise.
Values closer to $1$ show better performance.
\begin{definition}
Let $x_1,..,x_n$ be the ranks of the used premises and $y_1,..,y_m$ be the ranks of the unused premises.
Then, the AUC is defined as
$$ \text{AUC} = \frac{\sum_i^n\sum_j^m 1_{x_i > y_j}}{mn} $$
where $1_{x_i > y_j} = 1$ iff ${x_i > y_j}$ and zero otherwise.
\end{definition}

\section{Results}\label{sec:results}
For the evaluation we first topologically sort all the available \corn
facts using their proof dependencies, and then for each of these facts
we try to predict its proof dependencies by learning on all the
previous (in the topological sorting) facts and their
proofs.
As usual in such large-theory evaluations,
we thus emulate the standard situation of using AI/ATP assistance in
interactive theorem proving: For each conjecture $C$ we assume that
all formulas stated earlier in the development can be used to prove
$C$, and that all the proofs of previously stated theorems are
available for learning.\footnote{The standard (randomized)
  10-fold cross-validation %
 would in general
  allow to prove a conjecture using a theorem that was only proved
  later, resulting in cyclic proofs.}

Table~\ref{t1} shows the results. While the performance of
k-NN and naive Bayes is quite comparable to similar early evaluations
done for other systems and libraries, we get comparatively low
performance of MePo, which does not correspond to its good behavior in
our experiments with other systems. This could mean that the relations
between symbols in the \coq propositions differ quite significantly from
such relations in the other systems, and MePo-style heuristics need
some modifications for \coq. The best (Ensemble) performance means
that on average 75\% of the dependencies needed in the actual \coq
proof will be present among the first 100 hints recommended by the
Ensemble predictor trained on the previous proofs.

\begin{table}
\centering
\begin{tabular}{lccccc}\toprule
- & 100Cover (\%) & 100Precision & Recall & Auc & Rank \\ \midrule
MePo & 17.0	& 0.01004	& 954.45 & 0.2020	& 817.21  \\
k-NN	   & 72.3	& 0.07466	 & 451.28	 & 0.8341	 & 173.45	 \\
NBayes   & 72.6	& 0.08089	 & 448.80	 & 0.8366	 & 171.06	 \\
Ensemble & 74.9 & 0.08093  & 428.70  & 0.8490  & 158.37  \\\bottomrule
\end{tabular}
\caption{Experimental Results}
\label{t1}
\end{table}
\section{Conclusion}\label{sec:concl}

As far as we know, this is the first large-scale evaluation of
learning proof dependencies done on a larger library written in the
\coq system. The overall results are quite encouraging, showing that
guessing of the relevant parts of the library necessary for proving a
new \coq conjecture is not significantly harder than in other ITP systems. 

Future work includes better features for learning as already mentioned
above, and obviously also research on good automated proof techniques for
the \coq logic which could use the guessed premises for finishing off
smaller proofs automatically. These could be either translations into
the formats and logics used by the currently strongest ATP systems, or
direct implementations of custom proof-search methods working directly
in the \coq logic.

\section{Acknowledgments}
Thanks to Herman Geuvers for his help with clarifying the \coq terminology.
This research was supported by Austrian Science Fund (FWF):P26201, and by NWO grant \emph{Knowledge-based Automated Reasoning}.

\bibliography{ate11}
\bibliographystyle{plain}

\end{document}